\documentclass[12pt]{article}
\usepackage{amssymb}
\makeatletter
\if@titlepage
  \renewcommand\maketitle{\begin{titlepage}%
  \let\footnotesize\small
  \let\footnoterule\relax
  \null\vfil
  \vskip 60\p@
  \begin{center}%
    {\Large \bfseries\@title \par}%
    \vskip 3em%
    {\normalsize
     \lineskip .75em%
      \begin{tabular}[t]{c}%
        \@author
      \end{tabular}\par}%
      \vskip 1.5em%
    {\normalsize \@date \par}
  \end{center}\par
  \@thanks
  \vfil\null
  \end{titlepage}%
  \setcounter{footnote}{0}%
  \let\thanks\relax\let\maketitle\relax
  \gdef\@thanks{}\gdef\@author{}\gdef\@title{}}
\else
\renewcommand\maketitle{\par
  \begingroup
    \renewcommand\thefootnote{\fnsymbol{footnote}}%
    \def\@makefnmark{\hbox to\z@{$\m@th^{\@thefnmark}$\hss}}%
    \long\def\@makefntext##1{\parindent 1em\noindent
            \hbox to1.8em{\hss$\m@th^{\@thefnmark}$}##1}%
    \if@twocolumn
      \ifnum \col@number=\@ne
        \@maketitle
      \else
        \twocolumn[\@maketitle]%
      \fi
    \else
      \newpage
      \global\@topnum\z@   
      \@maketitle
    \fi
    \thispagestyle{plain}\@thanks
  \endgroup
  \setcounter{footnote}{0}%
  \let\thanks\relax
  \let\maketitle\relax\let\@maketitle\relax
  \gdef\@thanks{}\gdef\@author{}\gdef\@title{}}
\def\@maketitle{%
  \newpage
  \null
  \vskip 1em%
  \begin{center}%
    {\Large \bfseries\@title \par}%
    \vskip 1.5em%
    {\normalsize
      \lineskip .5em%
      \begin{tabular}[t]{c}%
        \@author
      \end{tabular}\par}%
    \vskip 1em%
    {\normalsize \@date}%
  \end{center}%
  \par
  \vskip 1.5em}
\fi
\renewcommand\section{\@startsection{section}{1}{\z@}%
                                     {-3.25ex\@plus -1ex \@minus -.2ex}%
                                     {1.5ex \@plus .2ex}%
                                     {\reset@font\normalsize\bfseries}}
\renewcommand\subsection{\@startsection{subsection}{2}{\z@}%
                                    {3.25ex \@plus1ex \@minus.2ex}%
                                    {-1em}%
                                    {\reset@font\normalsize\bfseries}}
\@addtoreset{equation}{section}
\newcommand{\note}[1]{\raisebox{1ex}{{\footnotesize \sf #1}}}
\newcommand{\rnote}[1]{\raisebox{1ex}{{\hspace*{-3mm} \scriptsize\sf#1}}
                       \hspace*{-4mm}}
\def\hsp{\vspace*{-1.5ex}}
\makeatother
\def\be{ \begin{equation}}          \def\ee{ \end{equation}}
\def\ba{ \begin{eqnarray}}          \def\ea{ \end{eqnarray}}
\def\nn{\nonumber}                  

\def\C{\mathbb{C}} \def\Z{\mathbb{Z}}
\def\o{\otimes}                     \def\bz{\mbox{\bf z}}
\def\bN{{\sf \bf N}}
\def\V{{\cal V}}
\def\W{{\cal W}}   \def\M{{\cal M}}   \def\G{{\cal G}} 
\def\a{\alpha }          \def\b{\beta }       \def\c{\gamma }   
               
\def\k{\kappa}           
\newtheorem{theo}{Theorem}

\newtheorem{prop}[theo]{Proposition}

\title{Generalization of the  Knizhnik-Zamolodchikov-Equations}
\author{{\sc Anton Yu.\ Alekseev \rnote{3}, 
           \  Andreas Recknagel \rnote{1}
         ,}\\[1mm] 
         {\sc   Volker Schomerus 
\rnote{2} } \\[5mm] 
\note{1} Institut f\"ur Theoretische Physik,  
\\ ETH - H\"onggerberg, CH-8093 Z\"urich, Switzerland
\\[1mm]
\note{2} II. Institut f\"ur Theoretische Physik, Universit\"at Hamburg,
\\ Luruper Chaussee 149, D--22761 Hamburg, Germany
\\[1mm]
\note{3} 
Institute of Theoretical Physics, Uppsala University,  
\\ Box 803, S-75108 Uppsala, Sweden}

\date{September 15, 1996}
\begin{document}
\begin{titlepage}      \maketitle       \thispagestyle{empty}

\begin{abstract}
\noindent In this letter we introduce a generalization of the 
Knizhnik-Zamo\-lo\-dchi\-kov equations from affine Lie algebras 
to a wide class of conformal field theories (not necessarily 
rational). The new equations describe correlations functions of 
primary fields and of a finite number of their descendents. Our 
proposal is based on Nahm's concept of {\em small spaces}  which 
provide adequate substitutes for the lowest energy subspaces in modules 
of affine Lie algebras. We explain how to construct the first order 
differential equations and investigate properties of the associated 
connections, thereby preparing the grounds for an analysis of 
quantum symmetries. The general considerations are 
illustrated in examples of Virasoro minimal models. 
\end{abstract}
\vspace*{-18cm}
{\tt {DESY 96-208 \hfill ETH-TH/96-35}}\\
{\tt {ESI-389 (1996) \hfill UUITP-21/96}}
\vfill
\noindent\phantom{wwwx}{\small e-mail: }{\small\tt alekseev@teorfys.uu.se, 
anderl@itp.phys.ethz.ch,}\\  
\phantom{wwwx{\small e-mail: }}{\small\tt vschomer@x4u2.desy.de} 
\end{titlepage}

Correlation functions of chiral primary fields in the 
Wess-Zumino-Novikov-Witten (WZNW) model satisfy a system 
of first order differential equations known as 
Knizhnik-Zamolodchikov (KZ) equations \cite{KZ}. 
During the last decade, they have led to a number 
of deep insights. First of all, the 
KZ-equations admit a reformulation as horizontality condition
for a flat connection on a certain vector bundle of finite rank; 
the monodromy theory of this connection beautifully 
explains the (quasi) quantum group structure in the WZNW-model 
\cite{Dri}. Another great achievement was to find explicit 
integral representations for the solutions of KZ-equations \cite{intsol}. 
These formulas have revealed a link to the Bethe-Ansatz for 
Gaudin-models \cite{KZintsys}. In \cite{BerFeld}, it was shown how 
to generalize the KZ-equations to WZNW models on higher genus Riemann 
surfaces; there, the equations include differentiations with respect to 
the moduli of the surface. Let us finally mention that the 
KZ-equations allow for discretizations (difference equations) 
which preserve much of the structure of their differential 
counterparts, see \cite{FrRe} and subsequent works. 

It is not yet clear how much of this beautiful theory can be 
carried over to the generalized KZ-equation that we are  
about to discuss; in this letter, the generalized system of 
first order differential equations will be introduced from 
the flat connection point of view, therefore we are in principle
in a position to discuss monodromy theory straight away. We will, 
however, only state some first elements here and give a detailed 
treatment elsewhere \cite{ARS}. The main focus of the present paper is the 
formulation of the generalized KZ-equations itself, which is 
obtained by restricting  the Friedan-Shenker connection \cite{FrSh} 
to a certain finite-dimensional subbundle of the (infinite-dimensional)
bundle of conformal correlation functions.

In order to define those finite rank subbundles, we first need a more 
precise description of our framework. We consider a chiral $W$-algebra 
$\W$ with finitely many bosonic generating fields  
\be
     W^s(z) = \sum_{n=-\infty}^{\infty} W^s_n z^{-n-h_s}
\ee
of positive integer conformal dimension (spin) $h_s=s$; here, 
$z\in \C$ is the left movers' coordinate and  $W^s_n,\ 
n\in\Z$, are the Laurent modes of $W^s(z)$. We identify 
$W^2(z)$ (or one of the fields of dimension two, should there be 
several) with the energy momentum tensor $T(z)$, whose Laurent 
modes $L_n$ satisfy the Virasoro algebra
\be \label{Vir}
   [ L_n,L_m ]
   = (n-m)\, L_{n+m} + \frac{c}{12} (n^3-n)\,\delta_{n+m,0}
\ee
with central charge $c$, whereas all the other generators $W^s(z)$ 
are Virasoro primary, i.e.\ 
\be \label{prim}
   [ L_n, W^s_m ]
    = \left( n(h_s-1)  - m\right) W^s_{n+m}\ \    
\ee 
holds for all $n,m \in \Z$. $\W$ is then linearly generated 
by these fields, their derivatives, and (derivatives of) 
normal-ordered products -- see e.g.\ \cite{NBlu} for 
more details on $W$-algebras. For our purposes, it will also be 
sufficient to regard $\W$ as the universal enveloping algebra 
of the Laurent modes of the $W^s(z)$.  

The representation theory of $W$-algebras, i.e.\ the classification of 
irreducible highest weight representations, can in principle be studied 
in a straightforward manner, although complete results are as yet 
available only for affine Lie algebras, for the Virasoro algebra 
and for some extensions and reductions thereof. On the other hand, to 
define and analyse the {\em fusion} of such representations is more 
difficult. For a large class of theories, namely the so-called 
{\em quasi-rational} CFTs, see below, Nahm has succeeded in giving 
a careful definition and in reducing the computation of fusion rules  
to a problem in finite-dimensional linear algebra \cite{Nah}.  
There is hope that by these methods  one can 
also find solutions of the associated Moore-Seiberg 
polynomial equations, but at present there are no simple 
algorithms which would allow to decipher information on {\em braiding} 
within this purely representation theoretic approach. We will see, 
on the other hand,  that the relevant data are encoded in the flat 
connection of the generalized KZ-equation. 

The cornerstone of the procedure in \cite{Nah} is the discovery that 
the irreducible representations of a quasi-rational CFT contain 
finite-dimen\-sio\-nal subspaces with special properties, and it is 
those subspaces that our formulation of generalized KZ-equations 
is based upon. 
Given a $W$-algebra $\W$, one may introduce the subspace (or sub-Lie
algebra) $\W_{--}$ spanned by all the modes $W_n$ with $n\leq -h(W)$,  
where $h(W)$ is the dimension of the field $W(z)$. Each irreducible 
highest weight module $\V$ of $\W$ contains a subspace 
$\V^a := \W_{--} \V$, and $\V$ is called {\em quasi-rational}  
if $n_V := \dim (\V/\V^a) < \infty$. In that case, we can choose 
a subspace $V$ of $\V$, called {\em small space}, such that 
$\dim V = n_V$ and $V+ \V^a$ is dense in $\V$. Note that while the 
integer $n_V$ is an {\em invariant} of the representation $\V$ of $\W$, 
there is some freedom of choice in selecting a  small space. 
Often, however, there are natural choices; in particular, one can 
arrange $V$ such that it has a basis of $L_0$-eigenvectors. 
In the following, all $W$-modules that occur are assumed to be 
quasi-rational modules, with small spaces carrying an $L_0$-grading. 

A CFT with finitely generated (bosonic) $W$-algebra is called 
{\em quasi-rational} if all the irreducible modules involved are 
quasi-rational. This class of CFTs includes all 
rational CFTs (see \cite{GaGo} for a proof) but in addition important 
non-rational cases. Among 
those, the superconformal CFTs associated to Calabi-Yau targets in 
string theory might be the most interesting examples. 
\footnote{Note that the restriction to bosonic $W$-algebras 
has merely been imposed
for simplicity and is not essential; in \cite{Matt} fermionic sectors
are discussed as well.}

We gain some first insight into the construction of small 
spaces by testing it in a standard example: the affine 
Lie-algebras $\hat \G_k$. The corresponding $W$-algebra is 
generated by currents $J^a(z)$,  $a = 1, \dots, \dim \G$,   
which are primary fields of dimension $h = 1$;  
the Virasoro modes are given by the Sugawara construction 
($h^\vee$ denotes the dual Coxeter number of $\G$)  
\be L_m = \frac{1}{k+h^\vee} \sum_{n \geq 0} : J^a_{m-n} 
     J_n^a : \ \ . \label{Ln} \ee
Irreducible highest weight representations $\V_\lambda$ of 
$\hat \G_k$ are induced from highest weight representations 
$\lambda$ of the zero mode algebra, which is isomorphic to the 
Lie algebra $\G$ itself. Since $\W_{--}$ is generated 
by the modes $J^a_n,\ n \leq -1$, $\V^a_\lambda$ 
contains all of $\V_\lambda$ except for the states of lowest energy. 
Consequently, a natural choice for the 
small space $V_\lambda$ of an irreducible highest weight 
representation of affine Lie algebras is to take this lowest 
energy subspace. As we will see later, the small spaces in 
general contain descendant states of higher energy as well; on the 
other hand, one may always use `singular vector relations' as  
in eq.\ (\ref{Ln}) in order to find an explicit basis. 

The most important abstract feature of small spaces realized in 
\cite{Nah} is their behaviour under the `fusion product 
action' of the $W$-algebra. Here, we can avoid all 
subtleties connected with a precise definition of this action, 
since only part of this structure is needed for our 
purposes: We consider tensor products of 
$N$ carrier spaces $\V_i$ of quasi-rational representations. 
$\V_\bN := \V_1 \o \dots \o \V_N$ 
admits $N$ pairwise commuting actions of the $W$-algebra $\W$, and 
we write $W^{(i)}_n$ for a mode $W_n$ acting on the 
$i^{th}$ tensor factor of $\V_\bN$. Now fix some 
tuple $\bz$ of complex parameters $z_i,\ i = 1, \dots, N,$ with 
$z_i \neq z_j$ for $i \neq j$,  and define the operators 
\be
 M_i(W_{n-h}) := W^{(i)}_{n-h} - (-1)^n \sum_{
     \stackrel{\scriptstyle j \neq i}{\scriptstyle k \geq 1}} 
 {k-1-n \choose k-1} (z_i - z_j)^{n-k} W^{(j)}_{k-h}
\label{miw}\ee     
for all $n \leq 0$ and all fields $W(z)\in\W$, where $h = h_W$ is 
the dimension of $W(z)$. Denote the universal enveloping algebra
of all the operators $M_i(W_{m})(\bz)$ by  $\M (\bz)$. This algebra 
could be regarded as $\W_{--}$ acting on $\V_\bN$
through the `fusion product representation'; 
then the elements of $\M (\bz)$ would be obtained from 
contour integration of field $W(z)$ with meromorphic 1-forms 
which are regular on $\C \setminus \bz$ and vanish at infinity.

Using the basic commutation relations (\ref{Vir}, \ref{prim}), 
one derives the formula  
$$
    \partial_j M_i(W_{n-h}) + [L_{-1}^{(j)}, M_i(W_{n-h})] 
     = - (n-1) \delta_{i,j} M_i (W_{n-1-h}) \ \ , 
$$
and since $\partial_i$ and $[L_{-1}^{(i)}, \cdot\, ]$ 
act as derivations on the family of algebras $ \M(\bz)$, 
we conclude that 
\be    \partial_i M + [L_{-1}^{(i)}, M ] \ \in \ \M(\bz) 
\label{cov} \ee
for all elements $M \in \M(\bz)$. Relation (\ref{cov}) is referred 
to as {\em covariance law} of the family $\M(\bz)$. 
 
Equipped with the algebras $\M(\bz)$, we may reformulate the main 
theorem on small spaces given in \cite{Nah}, which can be proven 
by a relatively simple recursive argument: 

Let $\Phi: \V_\bN \rightarrow S$ be a linear map into 
some vector space $S$. Suppose that \hsp
\begin{enumerate}
\item $\Phi$ annihilates all elements $M \in \M(\bz)$, i.e.\  
      that $\Phi M v = 0 $ for all $M \in \M(\bz)$ and all $v\in\V_\bN$;\hsp 
\item $\Phi$ vanishes on the $N$-fold tensor product $V_\bN \equiv V_1 
      \o \dots \o V_N \subset \V_\bN$ of small spaces. \hsp
\end{enumerate}
Then $\Phi$ vanishes on the full tensor product $\V_\bN$, i.e.\  
$\Phi \equiv 0$. 

This means that maps vanishing on $\M(\bz)$ 
are completely determined as soon as we know their action on the tensor
product of small spaces. We are interested in the following
consequence:

\begin{prop} 
Let $\V_\bN^*$ denote the dual of $\V_\bN$ and define 
$E_\bN(\bz) \subset \V_\bN^*$ to be the subspace of 
all linear forms on $\V_\bN$ which annihilate elements $M 
\in \M(\bz)$. Then there exists an injection of $E_\bN(\bz)$ 
into the tensor product $V_\bN$ of small spaces. 
\end{prop}

If this injection fails to be surjective, loosely speaking if the 
intersection $V_\bN \cap \M(\bz)\V_\bN$ is non-empty, we say that  
$V_\bN$ contains {\em spurious states} \cite{Nah}. 

Using the fusion product representation of $\W$ on $\V_\bN$, one 
would obtain the following immediate application of this proposition:  
Consider a tensor product of two modules $\V_{\alpha}$ and 
$\V_{\beta}$ 
and assume that the fusion product action of $\W$  on $\V_\a\otimes\V_\b$
can be decomposed into a sum of irreducible representations, i.e.\ 
$\V_\a\otimes\V_\b\cong \bigoplus_\c \bigoplus\! 
\raisebox{-4pt}{$\stackrel{\scriptstyle N_{\a\b}^\c}{\!\scriptstyle 1}$}\,\,
\V_\c$ as $\W$-modules; then 
the small space dimensions $n_\a$ and the fusion rules $N_{\a\b}^\c$ satisfy 
the inequality \cite{Nah}
$$
n_\a\, n_\b \geq \sum_\c N_{\a\b}^\c\, n_\c\ \ .
$$
In this situation, the inequality is strict if and only if 
$V_\a\otimes V_\b$ contains spurious states. 

Having collected these statements,  we are prepared to begin our construction 
of the generalized KZ-equations. We allow the coordinates 
$z_i$ to sweep out an open subset 
$U$ of the configuration space. A map $F: U \rightarrow 
\V_\bN^*$ is called holomorphic if the functions 
$\langle F(\bz)| v\rangle$ are holomorphic for all $v \in \V_\bN$. 
The trivial infinite rank vector bundle $U \times \V_\bN^*$ can be 
equipped with the following flat Friedan-Shenker (FS) connection 
(defined on holomorphic sections $F(\bz)$),  
\ba   \label{Conn}
  \nabla_\bN^{FS} &=& \sum_{i = 1}^N dz_i \nabla_i^{FS}  \ \ 
   \mbox{with}\\
  \nabla_i^{FS} F(\bz) &=& \partial_i F(\bz) -  F(\bz) 
  L_{-1}^{(i)}\ \ , \nn
\ea
where the last term is understood as composition of linear maps on 
$\V_\bN$. Flatness of the connection $\nabla_\bN^{FS}$ holds because 
of the commutation relation $[L_{-1}^{(i)}, L_{-1}^{(j)}] = 0$ 
and because $L_{-1}^{(i)}$ does not depend on the $z_j$. 

The crucial observation for our purposes is that $\nabla_{\bN}^{FS}$ 
restricts to a (flat) connection on a certain finite rank subbundle. 
Let us define the {\em space of generalized conformal blocks} as 
spanned by all holomorphic $F(\bz)$ which vanish on $\M(\bz)$, 
i.e.\ $ F(\bz) \ \in \ E_{\bN}(\bz)$. For such $F(\bz)$, we evaluate
$ \nabla^{FS}_i F(\bz) $ on $\M(\bz)$ and find
\ba
\langle \nabla^{FS}_i F(\bz)| M(\bz)v \rangle
& = & \  \ \langle \partial_i F(\bz)| M(\bz)v\rangle - 
   \langle F(\bz)| \ L_{-1}^{(i)}\  M(\bz)\  v \rangle \nn \\[1mm]
& = &  - \langle  F(\bz)| \partial_i M(\bz)v\rangle - 
   \langle F(\bz)| [ L_{-1}^{(i)},  M(\bz)] v \rangle \ \ . \nn 
\ea
In the last step we have dropped two terms which are zero due
to our assumption on $F(\bz)$. The covariance law (\ref{cov})
for the family of algebras $\M(\bz)$ finally gives 
$$  \langle F(\bz)|\M(\bz)v \rangle  = 0  \ \ \Rightarrow \ \ 
\langle \nabla_i^{FS} F(\bz)| M(\bz)v \rangle = 0 $$ 
for all $M(\bz)  \in \M(\bz)$ and $v \in \V_\bN$. When 
combined with Proposition 1 above, the result may be expressed 
as follows:
 
\begin{prop} {\em (Generalized KZ--connection)} Suppose
that the irreducible highest weight modules $\V_i$ 
are quasi-rational. Then the Friedan-Shenker connection 
(\ref{Conn}) descends to a flat connection 
\ba \nabla_\bN & = & \sum_i\, dz_i\, \nabla_i  \ \ 
   \mbox{ with }   \\[1mm] 
 \nabla_i &\equiv &  \partial_i - A_i  = \nabla_i^{FS}
   \mid_{\cal E} \nn
\ea
on the finite rank vector bundle ${\cal E}$ of generalized  
conformal blocks. The fibres $E_\bN(\bz)$ of ${\cal E}$ can 
be embedded into the tensor product $V_\bN$ of small spaces 
$V_i \subset \V_i$. 
\end{prop}    
      
The {\em generalized KZ-equation} is then simply the horizontality 
condition 
\be    \nabla_\bN F(\bz) = 
     {\textstyle \sum_i}\, dz_i\,(\partial_i - A_i)\,F(\bz) =  
     0 \ \ . \label{KZ} \ee 
At least if there are no spurious states, it is relatively 
easy to obtain explicit formulas for the connection. We 
recall this for the case of affine Lie algebras first and 
then give some new examples. 

We want to determine a formula for $\nabla_i F(\bz)$
when $F(\bz)$ vanishes on $\M(\bz)$. For these $F(\bz)$,  
$\nabla_i F(\bz)$ vanishes on $\M(\bz)$ too and hence it 
suffices to evaluate the form $\nabla_i F(\bz)$ on states 
in the tensor product of small spaces, i.e.\ to compute 
$$ \langle \nabla_i F(\bz)| v \rangle \ \ \mbox{ for all } 
 \ \ v \in V_\bN \ \ . $$
In the case of affine Lie algebras, we can exploit the fact that 
all the operators $J^{a(i)}_n$ with $n \geq 1$ vanish on 
$V_\bN$. With the help of eq.\ (\ref{Ln}) we find that 
\ba 
\lefteqn{ \langle \nabla_i F (\bz) , v \rangle}\nn \\[1mm] 
   & = & \nn
   \partial_i \langle F(\bz)|v\rangle - 
   \frac{1}{k + h ^{\vee}} \sum_a \ \langle F(\bz)|J^{a(i)}_{-1} 
    J^{a(i)}_0 v \rangle \\[1mm]
    \nn & = &       
   \partial_i \langle F(\bz)|v \rangle - 
   \frac{1}{k + h^{\vee}} \sum_{a,j \neq i} 
    {\frac{1}{z_i-z_j}} 
    \langle F(\bz)| J^{a(j)}_{0} J^{a(i)}_0 v \rangle\ \ .  \nn 
 \ea
When the operators $J^{a(i)}_0$ are replaced by their representation
matrices $t^a_i$ on the highest weight space $V_i$, our short computation 
produces the usual KZ-connection. Even though it is rather standard 
and easy to follow, we would like to describe the individual steps once 
more to distinguish particular properties of the example from generic 
features of the calculation. We start the evaluation of $\nabla_i$ 
by inserting its definition. This produces an operator $L_{-1}^{(i)}$ 
acting on the vector $v$ in the tensor product $V_\bN$ of small (or 
highest weight) spaces. The result is no longer in $V_\bN$ and  
can be expressed as a sum of terms where generators of $\W_{--}$ act 
on $V_\bN$, or -- to be more specific -- where $J_{-1}^{a(i)}$ act on 
$J_0^{a(i)} v \in V_\bN$. From the definition of our operators 
$M_i(W_n)$ and the requirement that $\langle F(\bz)|$ vanishes on 
$\M(\bz)$ we obtain 
\be 
 \langle F(\bz)| W^{(i)}_{n-h} = \sum_{\stackrel{\scriptstyle j\neq i}
{\scriptstyle k \geq 1}} 
 (-1)^n {k-1-n \choose k-1} (z_i - z_j)^{n-k}  
 \langle F(\bz)| W^{(j)}_{k-h}
\label{res}\ee
which is used to {\em reshuffle} the generators of $\W_{--}$ 
appearing on the lhs.  In our case, where $h=1$, it serves to 
replace operators $J_{-1}^{a(i)}$
by an infinite sum of $J^{a(j)}_{m}$ with $m \geq 0$. Since all 
$J^{a(j)}_{m}$ with $m > 0$ vanish on the lowest energy subspace, 
only the first terms  $J^{a(j)}_0$ act non-trivially on $V_\bN$ 
and contribute to the connection. 

Most of this analysis applies to more general examples of $W$-algebras, 
too. Only the discussion of the infinite sum in the end is based on 
some exceptional 
features of the affine Lie algebras and their representation 
theory. In general, a larger (but finite) number of terms in the sum may 
contribute because $h$ may be larger than 1 and because $W_n^{(j)} v$, 
$v \in V_\bN$, need not vanish for $n > 0$. 
Moreover, the action of operators $W_n^{(j)}$
can result in vectors outside of $V_\bN$ so that iterated 
reshuffling with the help of eq.\ (\ref{res}) is necessary. This 
means that an explicit formula for the generalized KZ-connection will 
have more terms than in the WZNW case -- the strategy for its derivation, 
however, remains unchanged.  
 
The degenerate models of the Virasoro algebra already display   
most aspects of the generic scenario. Their $W$-algebra is generated by the 
Virasoro field $T(z)$ with central charge $c = 1 - 6(p-q)^2/pq$, and  
their highest weight modules are induced from vectors $|h \rangle$ with 
$$      L_0 |h \rangle = h |h \rangle\ \ 
       , \ \ L_n |h \rangle = 0 $$ 
for all $n > 0 $. For $h = h(r,s) = ((pr-qs)^2 - (p-q)^2)/4pq$
with $p,q$ real and $r,s$ integer, those modules are quasi-rational
and the dimension of the small space $V = V^{p,q}_{r,s}$ 
can be computed to be    
$$   n \equiv n^{p,q}_{r,s} :=  \dim (V^{p,q}_{r,s}) = 
      \left\{ \begin{array}{l} {\rm min}\{ rs, (p-s)(q-r)\}   
      \hspace*{1em} \mbox{if $p,q$ integer}\ ,   \\[1mm]
             rs \hspace*{1em}\mbox{else}\ . \end{array}\right. $$
This number is just the energy level where
the first singular vector in the Verma module occurs: Observe 
that all elements $L_n, n \leq -2$, are in $\W_{--}$ so that 
we can choose a small space $V$ spanned by the states $L_{-1}^\nu
| h \rangle$ with $\nu = 0, 1, \dots $; when $\nu = n $, we 
find relations of the form  
$$L^n_{-1}|h\rangle  + |v'\rangle = 0$$
with some vector $|v'\rangle \in \V^a$,  see e.g.\ \cite{FeFu}, 
which shows that $\dim V \leq n$; a closer look at the 
structure of the normal ordered products \cite{NBlu} tells  
that indeed none of the vectors $L^\nu_{-1}|h\rangle$, 
$\nu=0,\ldots, n-1$, is contained in $\V^a$.            

Let us now restrict to the case $(r,s) = (2,1)$ for which one 
has 
\be    (L_{-1}^2 - \a L_{-2})|h(2,1) \rangle =0\ \  \mbox{with}
  \ \ \a  = \frac{2(2h(2,1) +1)}{3}\ .  
\label{sive}\ee
Thus, each small space is two-dimensional and we choose the one with 
basis vectors $|h\rangle$, $L_{-1}|h\rangle$. Following the 
procedure outlined above for WZNW models, one may derive 
formulas for the generalized KZ-connec\-tion corresponding 
to $N$-fold tensor products of $\V_{2,1}$. We state an 
explicit result for $\nabla_i = \partial_i - A^N_i$ in 
the case $N = 2$, where we obtain the following 
two $2^2 \times 2^2$-matrices, evaluated in the basis 
$v_1 = |h \rangle \o |h\rangle,\ v_2 = |h\rangle \o L_{-1} |h\rangle,\  
v_3 = L_{-1}|h\rangle \o |h\rangle$ 
and $v_4 = L_{-1} |h\rangle \o L_{-1} |h\rangle$:
\ba
A^{2}_1(\bz)&=&\pmatrix{0&0&1&0\cr 
                     0&0&0&1\cr
    {\alpha h\over(z_1-z_2)^2}&{\alpha\over z_1-z_2}&0&0\cr      
  {(\alpha+2)\alpha h\over(z_1-z_2)^3}&{\alpha(h+1)\over(z_1-z_2)^2}
   &{-\alpha^2\over(z_1-z_2)^2}&0} \label{aisa}\\[2mm] 
 A^{2}_2(\bz)&=&\pmatrix{0&1&0&0\cr
    {\alpha h\over(z_2-z_1)^2}&0&{\alpha\over z_2-z_1}&0\cr      
     0&0&0&1\cr
   {(\alpha+2)\alpha h\over(z_2-z_1)^3}&{-\alpha^2\over(z_2-z_1)^2}
     &{\alpha(h+1)\over(z_2-z_1)^2}&0}     
\label{aisb}\ea
The null vector (\ref{sive}), which already allowed us to 
determine the dimension of the small space, is again essential 
during the (multiple) reshuffling in the computation of $A^2_i$, 
cf.\ the remarks on the WZNW calculation. Above, we have tacitly 
assumed that $p$ and $q$ are chosen in such a way that no 
spurious states occur, which otherwise (e.g.\ for $p/q=2/5$) 
would have to be taken care of in the reshuffling procedure.

At this stage, let us recall a more general aspect of singular 
vectors: It is well known that by 
conformal covariance their existence 
leads to higher order differential equations on the conformal blocks. 
Of course, those can be replaced by systems of 
first order differential equations -- a procedure which is, however, 
highly non-unique. From this point of view, the remarkable feature of 
the generalized KZ-equations presented here is that they 
provide a {\em canonical} first order system. 

We would also like to remark that for coset models and certain 
generalizations, first order differential equations
for conformal correlators  are known for some time \cite{GK}, \cite{H}.
It might well be that an  investigation of the relation between those 
results and our generalized KZ-equations proves to be useful for obtaining 
integral representation of the correlation functions in a general case.

Turning back to the explicit formulas (\ref{aisa},\ref{aisb}) given above, 
we detect an obvious difference to the ordinary 
KZ-connection:  The connection matrices $A^2_i$ contain higher 
order poles. The appearance of higher singularities is somewhat 
disturbing since we expect the term $\nabla_i$ in our generalized 
KZ-connection to determine monodromy properties of the solutions when 
the `particle' at $z_i$ is moved around the one at $z_j$.
Such monodromies can be read off from the simple poles of 
the connection if there are no poles of higher order. So 
how is it possible to determine the monodromy from the generalized 
KZ-connection? 

To state the answer we need some more notations. Denoting the 
$L_0$-highest weight of some $\W$-module $\V$ by $h$, we 
define a map $d: V \rightarrow V$ by restricting $L_0 - h$ to the small 
space $V$, i.e.\ $d = 
L_0|_V -h$. In this way,  $V$ is endowed with a ($\Z$-valued) 
{\em energy grading}. From $d$ one obtains $N$ 
maps $d_i$ on $V_\bN$ so that $d_i$ detects the energy grading 
in the $i^{th}$ factor of the tensor product $V_\bN$, i.e.\  
$d_i = d^{(i)}$. Whenever $d$ does not vanish, the 
holomorphic, matrix-valued functions  
\be     D_{ij}(\bz) := (z_i - z_j)^{d_i  + d_j} \ \  
        \label{gt} \ee
are non-trivial and can be used to prove the following 
result:

\begin{prop} 
Let $\nabla_{\bN}^{ij}$ be obtained from $\nabla_{\bN}$ by 
a holomorphic gauge transformation with $D_{ij}(z)$ 
(defined as in eq.\  (\ref{gt})), i.e.\  
$$
    \nabla^{ij}_\bN \equiv D_{ij}(\bz) \nabla_\bN 
     D_{ij}^{-1}(\bz) = {\textstyle \sum_k}\, dz_k\,\nabla_k^{ij}. 
$$
Then the term $\nabla^{ij}_i$ has at most simple poles in 
$(z_i - z_j)$. Moreover, the residue at these poles is 
independent of $N$. 
\end{prop}
    
In other words, one can simply read off the monodromy from the residue of 
the generalized connections after having performed an explicitly known 
gauge transformation. This residue, and hence the 
monodromy, is not influenced by the possible presence of 
`particles' at $z_k$ with $k \neq i,j$. 
 
Let us also note that the monodromy does not depend (except 
{}from a similarity transformation) on whether we move $z_i$ 
around $z_j$ or $z_j$ around $z_i$. A mathematical formulation 
of this property involves the permutation operators $ P_{ij}: 
V_\bN \rightarrow V_\bN$ which permute the $i^{th}$ and $j^{th}$ 
factor in the tensor product $V_\bN$ of small spaces. In addition, let 
us define maps $\k_{ij}$ which exchange the $i^{th}$ and $j^{th}$
coordinate so that 
$$      \k_{ij} (z_i) = z_j \ \ , \ \ \k_{ij} (dz_i) = dz_j
\ \ , \ \ \k_{ij} (\partial_i) = \partial_j $$
etc. Our next proposition states that the generalized KZ-connection 
is symmetric under the simultaneous  exchange of small spaces and
coordinates.
 
\begin{prop} The connection
$\nabla_\bN$ is symmetric in the sense that  
$$    P_{ij} \nabla_\bN P_{ij} = \kappa_{ij}(\nabla_\bN) \ \ $$
holds for all indices $i,j = 1, \dots, N$, where $P_{ij}$ and 
$\k_{ij}$ denote permutation of small spaces and coordinates,
respectively. 
\end{prop}

Here, we have just described the first elements of a full monodromy
theory for the generalized KZ-equations. Being flat and symmetric, 
the connection $\nabla_{\bN}$ determines a representation of the 
braid group $B_{\bN}$ on the tensor product $V_{\bN}$ of small 
spaces. For the ordinary KZ-equations, this monodromy representation 
was computed explicitly by Drinfel'd \cite{Dri}. He showed that 
it may be obtained from a certain quasi-Hopf algebra and compared 
the latter with the quantum universal enveloping algebras 
$U_q(\G)$. Some parts of his analysis generalize to our framework 
and admit the reconstruction of a quasi-Hopf algebra. 
An explicit formula for the universal $R$-matrix, for instance, 
is based on Proposition 3. We plan to 
discuss the monodromy theory in a forthcoming paper \cite{ARS}.   
Let us nevertheless remark that existence of the quasi-Hopf algebras 
is already guaranteed by abstract reconstruction theorems in 
\cite{Rec}. The latter, however, make no predictions about 
the carrier spaces for representations of the quasi-Hopf
algebra. Our analysis leads us to identify these carrier 
spaces with the small spaces of quasi-rational models and
produces a {\em canonical quasi-Hopf algebra}. 

\bigskip
\noindent It is a pleasure to thank W.\ Nahm for stimulating and 
helpful discussions. We also thank J.\ Fr\"ohlich, M.\ Gaberdiel,
K.\ Gawedkzi and J.\ Teschner for useful comments. Hospitality of the 
Erwin-Schr\"odinger-Institute, Vienna, where part of this work 
was done, is gratefully acknowledged. 

\newcommand{\sbibitem}[1]{\vspace*{-1.5ex} \bibitem{#1}}

\vfill  
\end{document}